\documentclass[prl,superscriptaddress,twocolumn,showpacs,preprintnumbers,amsmath,aps]{revtex4}
\usepackage{epsfig,graphicx,hyperref}
\usepackage{dcolumn}
\usepackage{bm}

\newcommand{\be}{\begin{equation}}
\newcommand{\ee}{\end{equation}}
\newcommand{\ba}{\begin{eqnarray}}
\newcommand{\ea}{\end{eqnarray}}
\renewcommand{\phi}{\varphi}

\begin{document} 
\title{Nonperturbative effect of attractive forces 
in viscous liquids}

\author{Ludovic Berthier}
\affiliation{Laboratoire des Collo{\"\i}des, Verres
 et Nanomat{\'e}riaux, Universit{\'e} Montpellier II and CNRS,
34095 Montpellier, France}

\author{Gilles Tarjus} 
\affiliation{LPTMC, CNRS-UMR7600, Universit\'e Pierre et Marie Curie, boîte 121, 4 Place Jussieu, 75252 Paris cedex 05, France} 

\date{\today}
\begin{abstract}
We study the role of the attractive intermolecular forces in the viscous regime of a simple glass-forming liquid by using computer simulations. To this end, we compare the structure and the dynamics of a standard Lennard-Jones glass-forming liquid model with and without the attractive tail of the interaction potentials. The viscous slowing down of the two systems are found to be quantitatively and qualitatively different over a broad density range, whereas the static pair correlations remain close. The common assumption that the behaviour of dense nonassociated liquids is determined by the short-ranged repulsive part of the intermolecular potentials dramatically breaks down for the relaxation in the viscous liquid regime.
\end{abstract}



\maketitle

Differentiating the respective roles of repulsive and attractive intermolecular forces in the properties of fluids and liquids has a long history in statistical mechanics~\cite{hansen}. The so-called `van der Waals picture of liquids'~\cite{wca,wca_review}, \textit{i.e.} the predominance of the short-ranged repulsive part of the intermolecular potentials in determining the structure of dense nonassociated liquids, has proved very fruitful for predicting the pair correlation functions and the thermodynamics. Although not as thoroughly 
tested~\cite{jonas,berne,kushick,szamel}, 
it has been useful for the dynamics as well~\cite{chandler}.

More recently, this picture, in which the attractive part of the interactions is treated as a mere cohesive background amenable to perturbative treatment, has been transposed to the viscous (supercooled) liquid regime. 
A number of approaches either suggest or take for granted that the structure and the dynamics of viscous glass-forming liquids are controlled by the short-ranged repulsive forces. Among them are studies based on the mode-coupling theory of glasses~\cite{poon}, 
the self-consistent phonon theory~\cite{wolynes},
model potential energy 
landscapes~\cite{shell-debenedetti}, as well as recent work 
focusing on the correlations between pressure and energy 
fluctuations~\cite{pedersen,dyre} and on the density scaling of 
the relaxation time~\cite{coslovich}. Predominance of the short-ranged repulsive forces is also shared by the `jamming scenario', 
which postulates that the physics of glasses and glass-forming liquids 
is controlled by a zero-temperature critical point characteristic of the 
jamming~\cite{ohern,Xu} or glass~\cite{berthier-witten} transition 
of spheres with finite-ranged repulsive interactions, 
with the longer-ranged attraction considered as a perturbation. 

However, and quite surprisingly, the central hypothesis underpinning
this large set of ideas has never been directly studied.
It therefore appears timely to assess the role of the 
attractive intermolecular forces in the viscous liquid regime.
Our central conclusion is that in the regime that
is probed experimentally, attractive forces have little effect on the structure
of the liquid, but affect their dynamics in a strong, likely nonperturbative, 
way.

In this work, we compare by means of Molecular Dynamics simulation the structure and the dynamics of a standard model of glass-forming liquid, the 
Kob-Andersen 80:20 binary Lennard-Jones mixture~\cite{KA} and its reduction to the purely repulsive part of the pair potentials proposed by Weeks, Chandler and Andersen (WCA). In the following, the former is denoted by `LJ' and the latter by `WCA'. 
The interatomic pair potential between species $\alpha$ and $\beta$, with $\alpha, \beta = A, B$ is given in the two systems by
\begin{equation}
v_{\alpha \beta}(r) = 4\epsilon_{\alpha \beta} 
\left[\left( \frac{\sigma_{\alpha \beta}}{r}\right)^{12}- \left( \frac{\sigma_{\alpha \beta}}{r}\right)^{6} + C_{\alpha \beta} \right], \; r \leq 
r_{\alpha \beta}^c,
\nonumber
\end{equation}
and is zero otherwise, where $r_{\alpha \beta}^c$ is equal to the position of the minimum of $v_{\alpha \beta}(r)$ for the WCA potential and 
to a conventional cutoff of $2.5 \sigma_{\alpha \beta}$ for the standard LJ 
model; $C_{\alpha \beta}$ is a constant such that $v_{\alpha \beta}(r_{\alpha \beta}^c) =0$. The simulations are performed in the $NVE$ ensemble (after equilibration at a chosen temperature) with $N=900-1300$ particles (depending on the
density) and with
periodic boundary conditions. A broad range of density has been considered with $\rho$ from $1.1$ to $1.8$. Lengths, temperatures and times are given in 
units of $\sigma_{AA}$, $\epsilon_{AA}/k_B$, and $(m \sigma_{AA}^2/ 48 \epsilon_{AA})^{1/2}$ respectively. In line with the WCA theory, the two liquid models are compared at the same $(\rho, T)$ state points. Their pressure then differs, with the attractive interaction roughly providing a 
negative background term, as we have directly checked by studying the
equations of states, $P=P(\rho,T)$.

\begin{figure}
\psfig{file=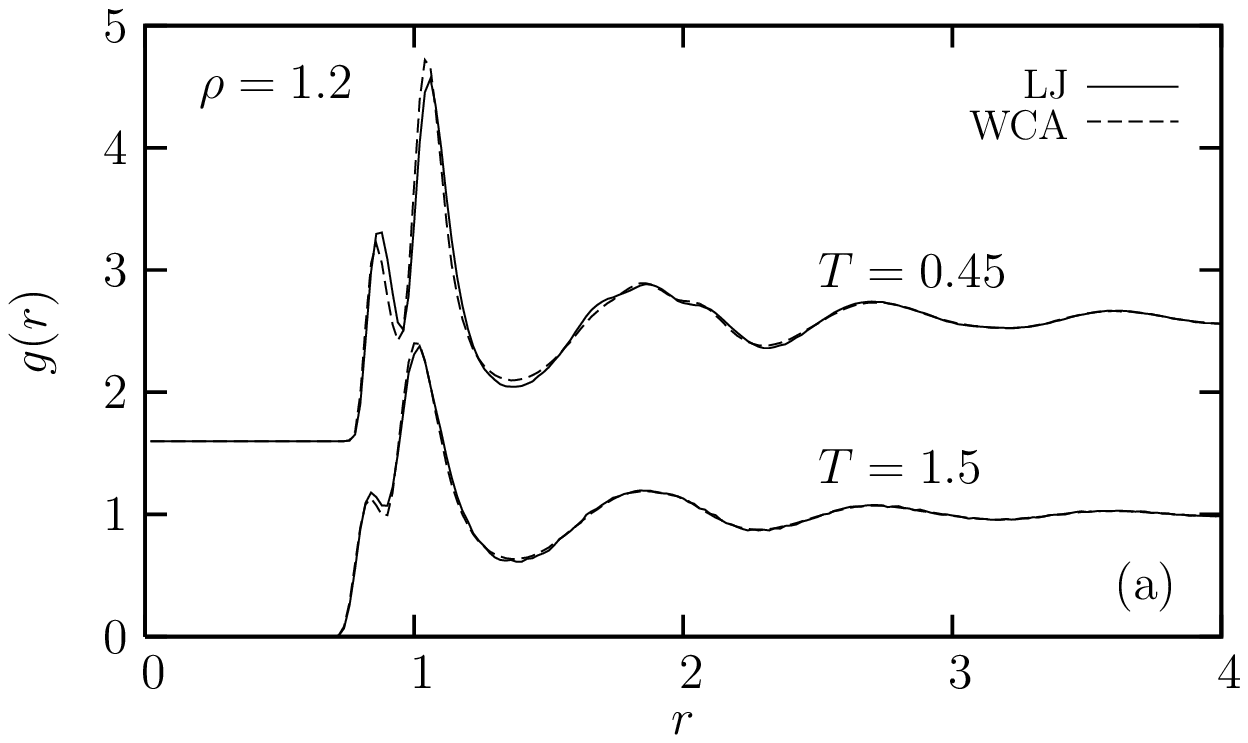,width=8.5cm}
\psfig{file=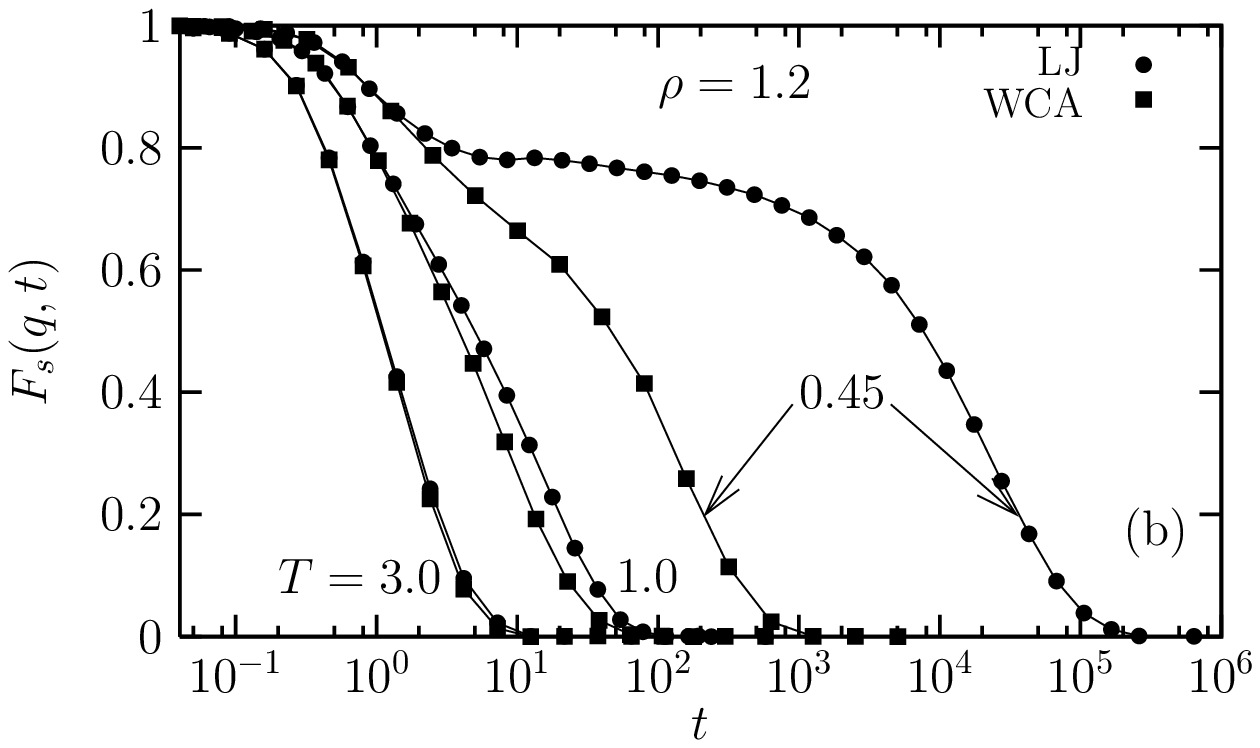,width=8.5cm}
\psfig{file=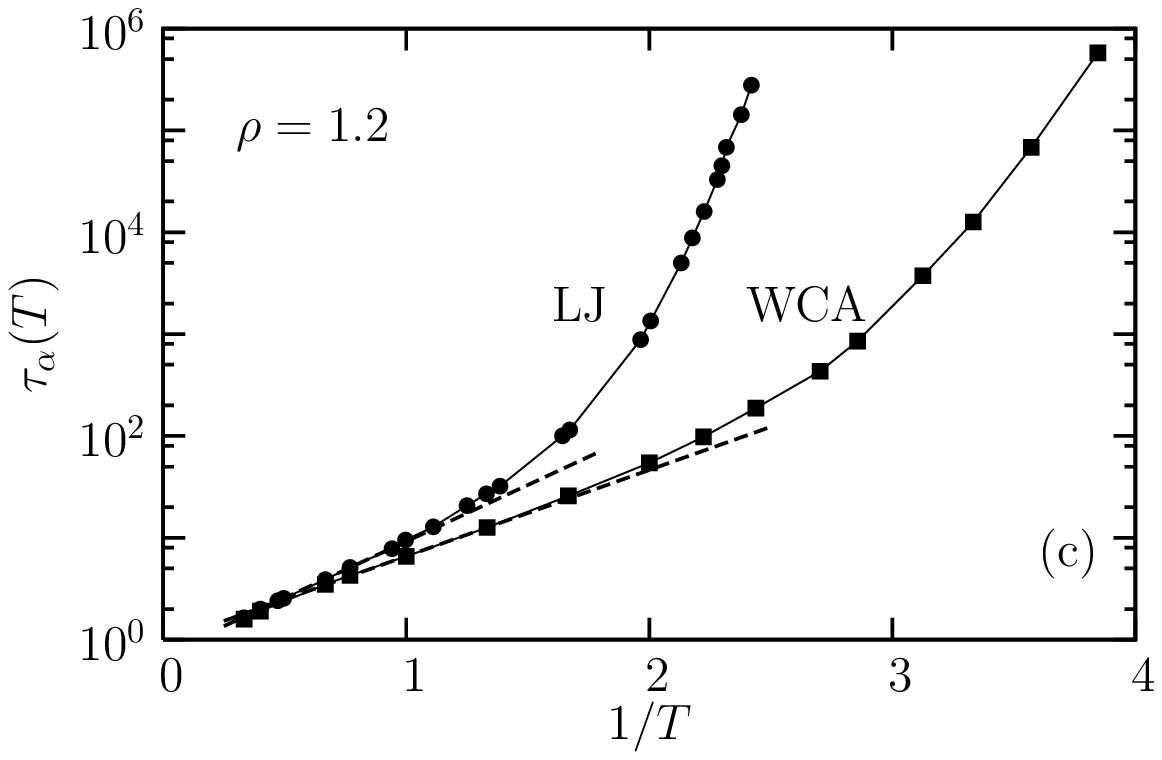,width=8.5cm}
\caption{\label{fs2} 
Comparison between the LJ and its WCA description at the typical 
liquid density $\rho=1.2$ for several temperatures. 
(a) Static pair correlation functions $g(r)$ 
as a function of $r$; 
(b) Time dependence of the self-intermediate scattering function 
for the majority component $A$;
(c) Arrhenius plot of the relaxation time $\tau_{\alpha}$ defined 
from $F_s(q,t=\tau_{\alpha})=1/e$ 
(the lowest $T$ is $0.43$ for LJ and $0.26$ for WCA), with 
high-$T$ Arrhenius fits shown as dashed lines.
Despite similar pair structure, both systems display dramatically different
dynamics.}
\end{figure}

We first consider the static structure of the liquids as characterized by the pair correlation functions $g_{\alpha \beta}(r)$. The pair correlation of the density fluctuations, $g(r)=\sum_{\alpha \beta}x_{\alpha}x_{\beta}g_{\alpha \beta}(r)$ with $x_{\alpha}$ the concentration of species $\alpha$, is displayed in
 Fig.~\ref{fs2} for two temperatures at the typical liquid density of $1.2$. As anticipated~\cite{wca}, the attractive forces play virtually no role in the high-$T$ liquid. The same is found at very high density ($\rho \gtrsim 1.6$, not shown here) for all $T$'s. We observe that the contribution of attraction remains very small over the whole range of $(\rho,T)$ under study. Inspection of the partial correlation functions $g_{\alpha \beta}(r)$ shows the same feature. A detectable effect is observed on the self-correlation of the minority component, $g_{BB}(r)$, but since $x_{B}=0.2$, it has little impact on $g(r)$.

Turning now to the dynamics, we present the data for the time dependence of the self intermediate scattering function \begin{equation}
F_s(q,t)=\frac{1}{N} \left\langle \sum_{j=1}^{N} e^{i \mathbf{q}.(\mathbf{r}_j(t)-\mathbf{r}_j(0))} \right\rangle,
\end{equation}
with $q \sigma_{AA} \simeq 7.2$,
which corresponds to the position of the 
the peak of the total static structure factor at $\rho=1.2$. 
We see that in the high-$T$ liquid (Fig.~\ref{fs2}b), the agreement between WCA and LJ is not perfect but is quite good (a factor of $2$ or less difference in the relaxation time for $T \gtrsim 1$, $\rho =1.2$). However, the difference between the two systems rapidly increases as one enters the viscous regime and becomes enormous at the lowest accessible temperatures. The relaxation time 
$\tau_{\alpha}$ of the WCA model is then more than $3$ orders of magnitude 
faster than that of the LJ model (see Figs.~\ref{fs2}b and \ref{fs2}c).

This is confirmed by other measures of the slowing down of relaxation. For instance, the empirically determined mode-coupling singularity temperature $T_c$, obtained by an algebraic fit of the temperature dependence of $\tau_{\alpha}$, is roughly divided by two when removing the attractive forces. We point out that, although not appreciated before, similar conclusions can be drawn by comparing the results already published in the literature for different Lennard-Jones 
mixtures~\cite{KA,lacevic} and for their 
WCA truncations~\cite{maibaum,castillo}.

We find that it is only by going to very high densities that the difference shrinks and the relaxation times of the two systems become more comparable. Yet, even at a density as high as $1.6$, \textit{i.e.} more than $30\%$ above the commonly used density of $1.2$, there is almost an order of magnitude difference between the relaxation times of the two systems, and at $\rho=1.8$, relaxation
times still differ by more than 20~\% at low temperatures, despite the fact
that the pair correlation functions virtually coincide at these densities. 
These observations prove that the dynamics is not primarily determined
by the static pair correlation functions, for instance raising some doubts on
the ability of the mode-coupling theory to properly describe the 
phenomenon.

The above results unambiguously show that the relaxation of the LJ is considerably slowed down by the presence of the attractive forces as the liquid is cooled down in the viscous regime and approaches the glass transition. One may however wonder whether this is a mere quantitative effect that can be accomodated by introducing an effective energy scale that takes into account the mean influence of the attractive part of the potentials by renormalizing the temperature 
scale~\cite{toxwaerd,coslovich}. To test this hypothesis, as well as to 
represent all of our data for different densities on the same graph, 
we have fitted the temperature dependence of the relaxation time in the high-$T$ regime to an Arrhenius formula: $\tau_{\alpha}\simeq \tau_{\infty}\exp(E_{\infty}(\rho)/T)$. This allows us to collapse the data at high $T$ 
with a good accuracy and to extract an effective activation energy scale 
$E_\infty(\rho)$. The latter can then be used to compare the relaxation data in the  presence and in the absence of attraction on a renormalized temperature scale $T/E_\infty(\rho)$ and therefore to test the above hypothesis. No emphasis is put on the physical meaning of this Arrhenius fit, 
which we take as a convenient and nonsingular representation 
of the high-$T$ data.

\begin{figure}
\psfig{file=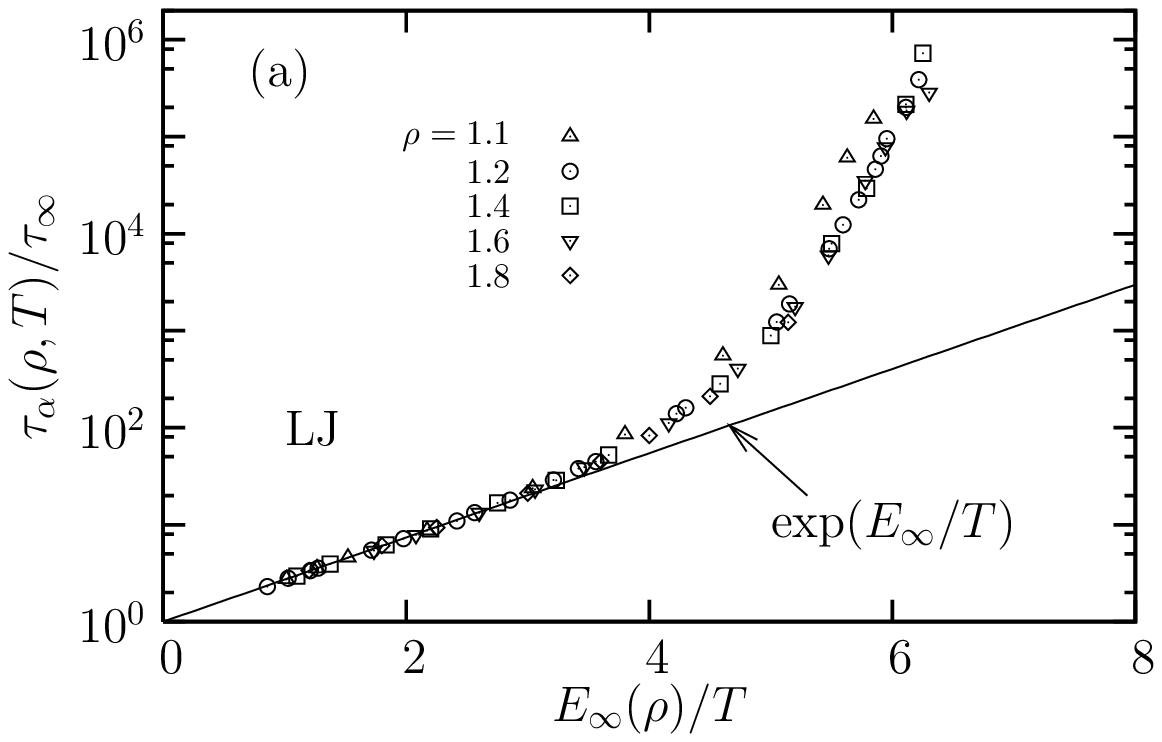,width=8.5cm}
\psfig{file=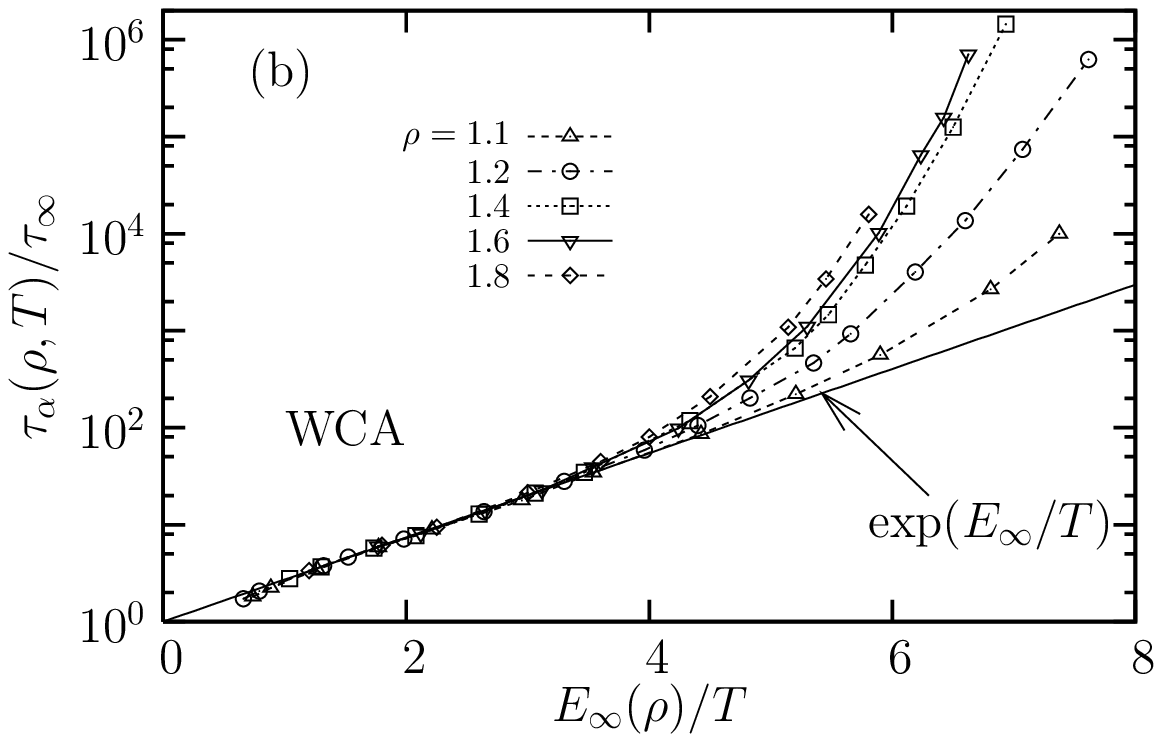,width=8.5cm}
\psfig{file=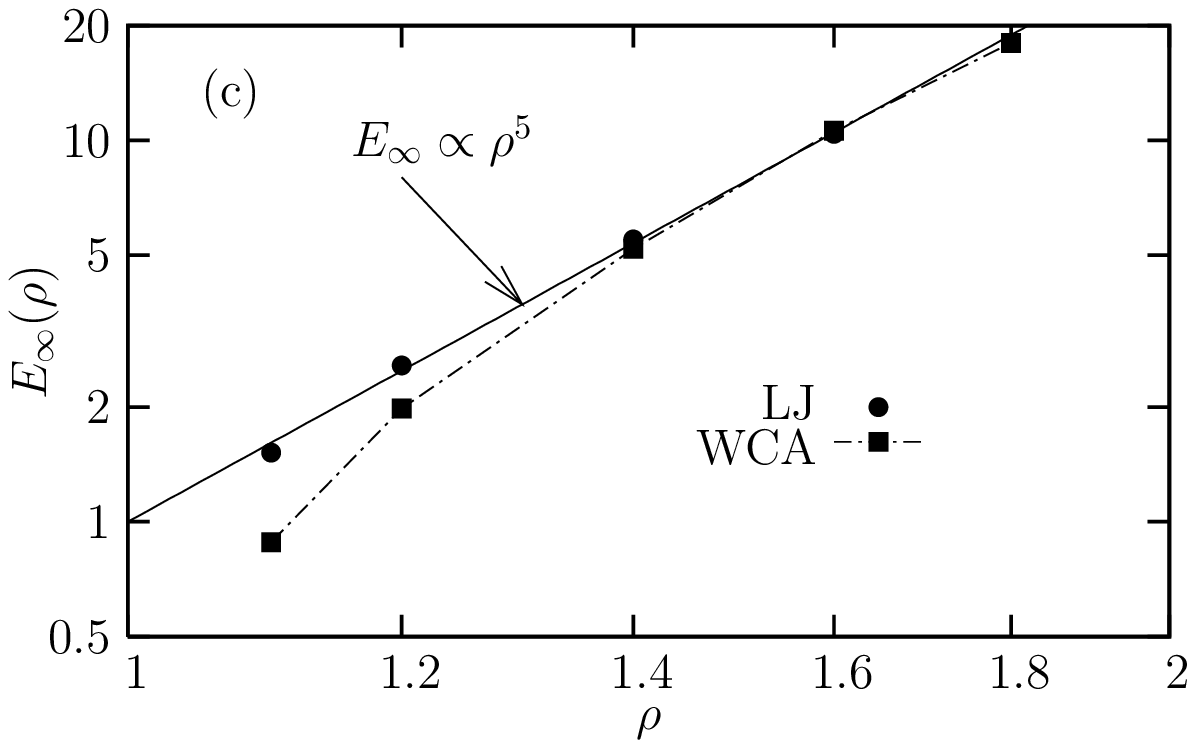,width=8.5cm}
\caption{\label{gilles} Rescaling of the relaxation data for the LJ and WCA
models over a wide range of densities and temperatures. We use the activation energy scale $E_\infty(\rho)$ obtained by fitting the high-$T$ data to an Arrhenius formula. (a) Arrhenius plot of the relaxation time for a scaled temperature $T/E_\infty(\rho)$ for LJ; 
(b) Same plot for WCA. 
Note the large change of fragility with density for
$\rho \lesssim 1.4$, not seen in the LJ data. 
(c) Density dependence of the activation energy scale for 
the two models, with LJ data fitted with a power law.}
\end{figure}

The results are shown in Figs.~\ref{gilles}a, \ref{gilles}b where we plot the logarithm of $\tau_{\alpha}$ for LJ and WCA for all densities between $1.1$ and $1.8$ as a function of the inverse of the scaled temperature, $E_\infty(\rho)/T$. By construction all curves coincide at high (scaled) temperature above some `onset',
$T/E_\infty(\rho)\simeq 0.3$, at which the viscous regime roughly
starts and departure from simple Arrhenius fit becomes significant. 
Below this onset temperature, we find that all LJ data essentially  
collapse onto a master curve (with a small deviation seen for the lowest density of $1.1$), as roughly do the WCA data for the three highest densities 
($\rho=1.4$,1.6, 1.8). The coincidence between LJ and WCA rescaled data is only fair at those densities, and the curves clearly diverge as one lowers the density to reach values more typical of regular supercooled liquids, \textit{i.e.} $\rho=1.2$ (compare Figs.~\ref{gilles}a and \ref{gilles}b).
The isochoric `fragility' of the WCA model is strongly 
density-dependent, which is reminiscent of the behaviour found 
in dense fluids of harmonic repulsive spheres~\cite{berthier-witten}.
This is clearly at variance with the almost constant 
isochoric fragility of the LJ. 

For completeness, we display in Fig.~\ref{gilles}c the density dependence of $E_\infty(\rho)$ for the 2 systems in a log-log representation. 
For LJ, it roughly goes as $\rho^5$, in agreement with previous work 
finding relaxation data collapse with the scaling 
variable $\rho^{\gamma}/T$ with $\gamma \simeq 5$~\cite{coslovich}.

The viscous slowdowns of the LJ and WCA models are therefore not only quantitatively different at a given density, they are also qualitatively distinct. 
The density scaling of the relaxation that is empirically found in real glassforming liquids and 
polymers~\cite{gilles4,gilles3,dreyfus,casalini1}, and in the LJ model as well (see Fig.~\ref{gilles}a and 
Refs.~\cite{gilles4,coslovich}) is strongly violated when attrative forces are truncated. These findings show that, contrary to expectations, the attractive components of the pair potentials play a crucial role in the viscous liquid regime when approaching the glass transition.
A purely repulsive WCA system of course displays a slowing down of relaxation that should end up in glass formation at low enough $T$, but some of the characteristics of this slowing down, including the absence of density scaling of the relaxation time, are at odds with the behavior of the full LJ model it is supposed to describe, and of real glass-formers.

\begin{figure}
\psfig{file=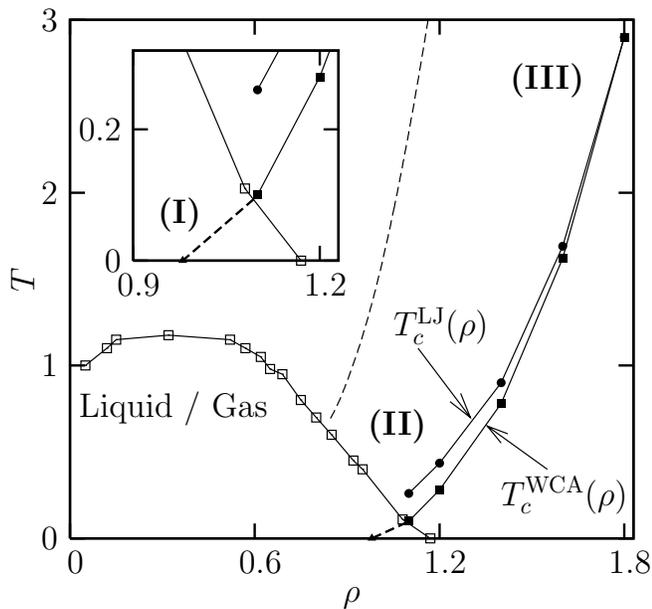,width=8.5cm}
\caption{\label{phase}
Phase diagram of the LJ model. The melting line (shown as a dashed line) 
is for the one-component model~\cite{depablo}. 
We distinguish $3$ regions. 
(I) is inside the liquid-gas spinodal and is only accessible 
for WCA: see the zoom in the inset.
The extrapolated curve shown as a dotted line 
ends in a putative zero-temperature 
singularity~\cite{ohern,berthier-witten}. 
(II) is the range corresponding to the experimentally accessible 
liquid regime, where large differences are found in 
the dynamics of the two models.
(III) is the high-$T$ and high-$\rho$ region only reachable 
in computer simulations of model systems, in which the role of 
the attractive forces becomes small or negligible.}
\end{figure}

Finally, we map out in Fig.~\ref{phase}
the various regimes studied here in a $(\rho, T)$ phase diagram and discuss the relevance of our findings to real glass-formers. 
On top of the thermodynamic transition lines, we have plotted the 
empirically determined mode-coupling line, $T_c(\rho)$, as an indication of the trend for the (isochronic) glass transition line in the diagram. One can schematically distinguish three regions. Region (I), inside the gas-liquid coexistence curve  (or spinodal~\cite{sri}), can only be accessed by removing the attractive part of the potentials. This is the region that could be controlled by a zero-temperature jamming~\cite{ohern,Xu} or glass~\cite{berthier-witten} critical 
point. Region (II) is the experimentally accessible range of glass-forming liquids, for which, typically, the pressure can be varied from 1 bar to 10 kbars with an associated density variation of $20-25\%$. Finally, region (III) is a high-$T$, high-$\rho$ regime that is only reachable in computer simulations of model systems. It is only in this region that the role of the attractive forces on both the statics and the dynamics essentially becomes negligible and could
be treated perturbatively.

The important result obtained in our study is that the dynamics in region (II) at temperatures characteristic of viscous liquid behavior is strongly influenced, and in a highly nontrivial and nonperturvative way, 
by the attractive forces. It seems therefore unlikely that an extrapolation from either regime (I) or regime (III) could fruitfully describe real glass-forming liquids in region (II). 
Thus, the common assumption that the behaviour of dense nonassociated 
liquids is determined by the short-ranged repulsive part of the intermolecular 
potentials  dramatically breaks down for the dynamics in the 
viscous liquid regime.
This striking result should motivate further work to reconsider
the interplay between structure and dynamics, density and temperature,
or else jamming and thermal activation in the glass formation.

\acknowledgments
We acknowledge partial support from the ANR Dynhet.

\end{document}